\documentclass[12pt, reqno]{amsart}

\usepackage{tikz}
\usetikzlibrary{positioning}
\usetikzlibrary{hobby,decorations.markings}
\usepackage{pgfplots}
\usepackage{float} 
\usepackage{tkz-graph} 
\usepackage{tkz-berge} 
\usepackage{mathtools} 
\usepackage{xfrac} 
\usepackage{amssymb} 

\makeatletter
\newcommand{\vast}{\bBigg@{4}}
\newcommand{\Vast}{\bBigg@{5}}
\newcommand{\vastt}{\bBigg@{6}}
\newcommand{\Vastt}{\bBigg@{7}}
\makeatother

\newcommand{\pn}[2]{P_n(#1) \, \mathbf{#2}}
\newcommand{\Pn}[2]{\mathcal{P}_n(#1) \, \mathbf{#2}}

\newtheorem{lemma}{Lemma}
\newtheorem{proposition}{Proposition}
\newtheorem{theorem}{\textit{Theorem}}[section]

\newtheorem{remark}{Remark}
\newtheorem{corollary}{Corollary}

\allowdisplaybreaks 

\usepackage{lineno}

\begin{document}

\title[Finding the minimum norm and center density]{Finding  the minimum norm and center density of cyclic lattices via nonlinear systems}


\author{William Lima da Silva Pinto}
\address{São Paulo State University - Brazil}
\curraddr{}
\email{william.lima@unesp.br}
\thanks{This work was supported by FAPESP Proc. 2019/20800-8 and 2013/25977-7. }

\author{Carina Alves}
\address{Department of Mathematics, São Paulo State University - Brazil}
\curraddr{}
\email{carina.alves@unesp.br}
\thanks{}


\date{}

\dedicatory{}

\begin{abstract}

Lattices with a circulant generator matrix represent a subclass of cyclic lattices. This subclass can be described by a basis containing a vector and its circular shifts. In this paper, we present certain conditions under which the norm expression of an arbitrary vector of this type of lattice is substantially simplified, and then investigate some of the lattices obtained under these conditions. 
We exhibit systems of nonlinear equations whose solutions yield lattices as dense as $D_n$ in odd dimensions. As far as even dimensions, we obtain lattices denser than $A_n$ as long as $n \in 2\mathbb{Z} \backslash 4\mathbb{Z}$.

\end{abstract}

\maketitle

\section{Introduction}
%

An $n$-dimensional lattice is a discrete additive subgroup of $\mathbb{R}^n$, consisting of linear combinations of linearly independent vectors in $\mathbb{R}^n$ with integer coefficients. We say it is a full rank lattice if the number of those linearly independent vectors is equal to the lattice dimension. Lattice properties are related to various areas, such as signal processing \cite{conway} , \cite{signal1} and cryptography \cite{cryptography1}, \cite{cryptography2}. The sphere packing problem aims to find out how dense a large number of identical spheres can be packed together in the Euclidean space. The  {packing density} of a lattice $\Lambda$ is the proportion of the space $\mathbb{R}^n$ covered by the non-overlapping spheres of maximum radius centered at the points of $\Lambda$ and can be obtained in terms of the minimum norm $\lvert \Lambda \rvert = \min \{ \Vert \textbf{x} \Vert^2 \colon \textbf{x} \in \Lambda, \, \textbf{x}\neq \mathbf{0} \}$.

 Lattices with high packing densities are usually associated with good signal constellations over Gaussian channels \cite{conway},\cite{2}.
The densest possible lattice packings have only been determined in dimensions $1$ to $8$ \cite{conway} and $24$ \cite{cohn}. In \cite{2}, rotated $n$-dimensional lattices (including $D_4,$ $K_{12}$ and $\Lambda_{16}$), good for both Gaussian and Rayleigh fading channels have been constructed. More recently, in \cite{agnaldo},  rotated $A_n-$lattices, for $n=2^{r-2}-1$, $r \geq 4$ have been proposed.
  If $G$ is the matrix determined by some basis of a full rank lattice $\Lambda$, that is, a generator matrix, then the packing density depends directly of the parameter $\delta (\Lambda)=(\sqrt{\lvert \Lambda \rvert}/2)^n/\lvert \det G\rvert$, called center density \cite{conway}. 
 However, it is generally not an easy task to compute $\lvert \Lambda \rvert$. In fact, the shortest vector problem (SVP) is an NP-hard problem in general \cite{nphard}, \cite{nphard2}, and has also drawn the attention of ma-\linebreak thematicians and computer scientists because of its relation with integer programming \cite{integer1}, \cite{integer2}.\

 Another lattice problem related to that is to determine the number of vectors of $\Lambda$ with minimum norm, which is known as the kissing number problem (KNP). The exact number is known for dimensions 1, 2, 3, 8, and 24 \cite{conway}, \cite{musin} but there exist bounds in many other dimensions, for example, \cite{survey}, \cite{maehara}.

Classes of lattices that have the calculation of $\lvert \Lambda \rvert$ simplified, either by construction \cite{cari} or by algorithms \cite{conway},\cite{svp}, \cite{svp2}, are much desired. In this paper we work around cyclic lattices, a particular class of lattices that is relatively good for that purpose, and was first addressed by Micciancio \cite{mici}. Cyclic lattices are those which applying a circular shift operator to one of its vectors will result in another vector from the same lattice. In other words, cyclic lattices are those that are closed under such operator.

In particular,  performing circular shifts over a vector $\mathbf{u}\in\mathbb{R}^n$ yields a basis for a cyclic lattice. 


A more common approach has been to assume $\mathbf{u}\in\mathbb{Z}^n$ as per \cite{fukshansky}.
In the present work, we study the general case, exhibiting some strategies to simplify the calculation of $\lvert \Lambda \rvert$ and increasing of $\delta(\Lambda)$ under certain conditions. We end up with nonlinear systems of equations whose solutions yield lattices as dense as $D_n$ in odd dimensions. 

This paper is organized as follows: in Section 2 we discuss cyclic lattices defined over circular shifts of an arbitrary vector and calculate the norm of an arbitrary vector through some properties of the inner product of a vector and its circular shifts. In Sections 3 and 4 we provide conditions under which the norm is simplified and further obtain lattices with good properties.

\section{Generalizing the Norm}
\noindent Let $n\geq 2$ and define the circular shift operator $\textit{rot}:\mathbb{R}^n\rightarrow \mathbb{R}^n$ by
\[
\textit{rot}(x_1,x_2,...,x_{n-1},x_n)=(x_n,x_1,x_2,...,x_{n-1}).
\]

A lattice $\Lambda$ is called cyclic if it is closed under \textit{rot}, that is, $\textit{rot} ( \Lambda) = \Lambda$. 

If there exists a vector $\mathbf{u}=(\rho_1,\rho_2,...,\rho_n)\in \mathbb{R}^n$ such that $\{ \mathbf{u}, \textit{rot}(\mathbf{u}),...,\linebreak \textit{rot}^{\,n-1} (\mathbf{u}) \}$ is a basis for $\Lambda$, then $\Lambda$ is evidently cyclic. We denote such lattice as $\Lambda_\mathbf{u}$. A lattice $\Lambda_\mathbf{u}$ has a circulant generator matrix \cite{circulant1}, \cite{circulant2} as follows:
\[
G_{\mathbf{u}}=
\begin{pmatrix}
\rho_1 & \rho_2 & ... & \rho_n \\
\rho_n & \rho_1 & ... & \rho_{n-1} \\
\vdots & \vdots & \ddots & \vdots \\
\rho_2 & \rho_3 & ... & \rho_1
\end{pmatrix}.
\]


Some general properties of such lattices have been discussed in \cite{fukshansky} when $\mathbf{u} \in \mathbb{Z}^n$. We shall investigate throughout this paper, however, the broader case $\mathbf{u}\in\mathbb{R}^n$ and by a different approach. We want conditions over $\mathbf{u}$ such that $\det G_\mathbf{u}\neq 0$ and $\Lambda_\mathbf{u}$ is as dense as possible.\

From now on, let $a,b\in\mathbb{R}$ be the coefficients that multiply $t^{n-1}$ and $t^{n-2}$ in $f(t)=\prod_{i=1}^n(t-\rho_i)\in\mathbb{R}[t]$, respectively. By the Vieta's formulas \cite{vinberg}, $-a=\sum_{i=1}^n\rho_i$ and $b=\sum_{i<j}\rho_i\rho_j$. Consequently, $\sum_{i=1}^n\rho_i^2=a^2-2b$.

Now, given an arbitrary vector $\textit{\textbf{w}} \in \Lambda_\mathbf{u}$, we are interested in computing $\Vert \textit{\textbf{w}} \Vert^2$, in order to investigate 
\begin{equation} \label{minimum}  \lvert \Lambda_\mathbf{u}  \rvert  =\min\{\Vert \textit{\textbf{w}}\Vert^2 \colon \textit{\textbf{w}} \in \Lambda_{\textbf{u}}, \textit{\textbf{w}} \neq 0\},\end{equation}
and the amount of minimal vectors
 of $\Lambda_\mathbf{u}$, which can be defined as  
 \begin{equation} \label{qminimum}
 \lvert S(\Lambda_\mathbf{u}) \rvert \vcentcolon = \# \{ \textit{\textbf{w}} \in \Lambda_\mathbf{u}  \colon \Vert \textit{\textbf{w}} \Vert^2 = \lvert \Lambda_\mathbf{u} \rvert \}.\end{equation} 
The number of minimal vectors is called kissing number and is often denoted by $\kappa$. 

For each $r\in \{ 1,2,...,n-1\}$ define $I_n = \{ 1, 2, \cdots, n \}$ and
\begin{equation}  \label{expre1} 
\pn{r}{x} =  \smashoperator{\sum_{\substack{i,j\in I_n \\ i<j \\ j-i=r}}}x_ix_j,
\qquad
\forall \mathbf{x}=(x_1,...,x_n)\in\mathbb{R}^n.
\end{equation} 

\begin{lemma} \label{lema1}
Let  $n\geq 2$ and $\mathbf{x}=(x_1,...,x_n)\in \mathbb{R}^n$. If 
 $0\leq k_1 < k_2 \leq n-1$, then
\[
\langle \textit{rot}^{k_1} (\mathbf{x}),\textit{rot}^{k_2}(\mathbf{x}) \rangle=\pn{k_2-k_1}{x}+\pn{n-(k_2-k_1)}{x}.
\]
\begin{proof}
Note that
\[
\textit{rot}^{k_2-k_1}(\mathbf{x})=(x_{n-(k_2-k_1-1)},x_{n-(k_2-k_1-2)},...,x_n,x_1,...,x_{n-(k_2-k_1)}).
\]
Hence,
\begin{align*}
\langle \textit{rot}^{k_1}(\mathbf{x}),\textit{rot}^{k_2}(\mathbf{x})\rangle = &\langle \mathbf{x},\textit{rot}^{k_2-k_1}(\mathbf{x})\rangle\\
=&\left({x_1 x_{n-(k_2-k_1)+1}+x_2 x_{n-(k_2-k_1)+2}+...+x_{k_2-k_1}x_n}\right) +\\ +& 
\left({x_{k_2-k_1+1}x_1+...+x_n x_{n-(k_2-k_1)}}\right) \\
 \stackrel{(\ref{expre1})}{=}&  \pn{n-(k_2-k_1)}{x} + \pn{k_2-k_1}{x},
\end{align*}
which proves the lemma.
\end{proof}
\end{lemma}

Inspired by the Lemma \ref{lema1}, for each $r\in \{ 1,2,...,n-1\}$ define
\begin{equation}  \label{usalema1}
\Pn{r}{x}=   \smashoperator{\sum_{\substack{i<j \\ j-i\in\{r,n-r\}}}}x_ix_j,
\qquad
\forall \mathbf{x}=(x_1,\cdots,x_n)\in\mathbb{R}^n.
\end{equation} 
Hence, for $0 \leq k_1 < k_2 \leq n - 1$,
\begin{align} 
    \label{condi1}
\pn{k_2-k_1}{x}+\pn{n-(k_2-k_1)}{x} &=
\begin{cases}
{2\Pn{k_2-k_1}{x},}&{\text{if}}\, k_2-k_1=\frac{n}{2}\\
{\Pn{k_2-k_1}{x},}&{\text{if}}\, k_2-k_1\neq\frac{n}{2}.
\end{cases} 
\end{align}


For example, 
 $\mathcal{P}_5(1) \, \mathbf{x}=x_1 x_2+x_2 x_3+ x_3 x_4+x_4 x_5 + x_1 x_5$ and $\mathcal{P}_5(2) \, \mathbf{x}=x_1 x_3+x_2 x_4+ x_3 x_5+x_1 x_4 + x_2 x_5$. Moreover, $b=\sum_{i<j} \rho_i \rho_j = \mathcal{P}_5(1) \, \mathbf{u} + \mathcal{P}_5(2) \, \mathbf{u}$. 




\begin{proposition}
Let $\rho_1,...,\rho_n\in\mathbb{R}$, $\mathbf{u}=(\rho_1,...,\rho_n)$ and $f(t)= \prod_{i=1}^n(t-\rho_i)$. If $b\in\mathbb{R}$ is the coefficient that multiplies $t^{n-2}$ in $f(t)$, then
\[
b=\sum_{r=1}^{\lfloor \frac{n}{2}\rfloor} \Pn{r}{u}.
\]
\begin{proof}
By Vieta's formulas \cite{vinberg}, if $\tau_n = (1+(-1)^n)/2	$, then
\begin{align*}
b&=\sum_{\substack{i,j\in I_n \\ i<j}}\rho_i\rho_j
=\sum_{r=1}^{n-1}\sum_{\substack{i,j\in I_n \\ i<j \\ j-i=r}}\rho_i\rho_j
\stackrel{(\ref{expre1})}{=}\sum_{r=1}^{n-1} \pn{r}{u}
=\tau_n\pn{\tfrac{n}{2}}{u}+ \sum_{\substack{r=1 \\ r\neq\frac{n}{2}}}^{n-1} \pn{r}{u},
\end{align*}
Hence, if $n$ is even,
\begin{align*}
b=& P_n\bigg( \frac{n}{2}\bigg) \, \mathbf{u}+\bigg( \pn{1}{u}+\pn{2}{u}+...+P_n\bigg(\frac{n}{2}-1 \bigg) \, \mathbf{u} +\\ +& P_n\bigg(\frac{n}{2}+1 \bigg)\, \mathbf{u}+ ...+\pn{n-1}{u} \bigg) \stackrel{(\ref{condi1})}{=}\mathcal{P}_n\bigg( \frac{n}{2}\bigg) \, \mathbf{u}+\bigg( \Pn{1}{u}+\\ +& \Pn{2}{u}+...+\mathcal{P}_n\bigg(\frac{n}{2}-1\bigg) \, \mathbf{u} \bigg)=\sum_{r=1}^{\frac{n}{2}}\Pn{r}{u}
=\sum_{r=1}^{\lfloor \frac{n}{2}\rfloor}\Pn{r}{u}.
\end{align*}
On the other hand, if $n$ is odd,
\begin{align*}
b&= \pn{1}{u}+\pn{2}{u}+...+\pn{n-1}{u}\\
&\stackrel{(\ref{condi1})}{=} \Pn{1}{u}+\Pn{2}{u}+...+\mathcal{P}_n\bigg(\frac{n-1}{2}\bigg) \, \mathbf{u} \\
&=\sum_{r=1}^{\frac{n-1}{2}}\Pn{r}{u}=\sum_{r=1}^{\lfloor \frac{n}{2}\rfloor}\Pn{r}{u},
\end{align*} which proves the proposition. \end{proof}
\end{proposition}

When we consider $\textit{\textbf{w}}\in \Lambda_{\textbf{u}} $ we can characterize $\Vert \textit{\textbf{w}} \Vert^2$ as in the following theorem.

\begin{theorem}\label{norma^2}
Let $n\geq 2$ and $\mathbf{u}=(\rho_1,...,\rho_n)\in \mathbb{R}^n$ such that $\det G_{\mathbf{u}}\neq 0$. If $ \textbf{w}=  \sum_{i=1}^n x_i \textit{rot}^{\,i-1}(\mathbf{u}) \in\Lambda_\mathbf{u}$,  then
\[
\Vert  \textit{\textbf{w}} \Vert^2=(a^2-2b)\sum_{i=1}^n x_i^2+2\sum_{r=1}^{\lfloor \frac{n-1}{2}\rfloor} \Pn{r}{u}\, \Pn{r}{x}+\tau_n\Big(4\Pn{\tfrac{n}{2}}{u} \, \Pn{\tfrac{n}{2}}{x}\Big),
\]
where  $(x_1,\cdots, x_n)\in\mathbb{Z}^n$,   $a,b\in\mathbb{R}$ are the coefficients multiplying $t^{n-1}$ and $t^{n-2}$ respectively, in $f(t)= \prod_{i=1}^{n}(t-\rho_i)$, and $\tau_n= (1+(-1)^n)/2$.

\begin{proof}
If  $\textit{\textbf{w}}=  \sum_{i=1}^n x_i \textit{rot}^{i-1}(\mathbf{u}) \in\Lambda_\mathbf{u}$ then
\begin{align*}
\Vert \textit{\textbf{w}} \Vert^2 &=\Bigg\langle \sum_{i=1}^n x_i \textit{rot}^{\,i-1}(\mathbf{u}) , \sum_{i=1}^n x_i \textit{rot}^{\,i-1}(\mathbf{u}) \Bigg\rangle 
=\sum_{i=1}^n \sum_{j=1}^n \langle x_i\textit{rot}^{\,i-1}(\mathbf{u}),\\ &  x_j\textit{rot}^{\,j-1}(\mathbf{u}) \rangle 
=\sum_{i=1}^n \sum_{j=1}^n x_i x_j \langle \textit{rot}^{\,i-1}(\mathbf{u}),  \textit{rot}^{\,j-1}(\mathbf{u}) \rangle \\
&=\sum_{i=1}^n x_i^2 \langle \textit{rot}^{i-1}(\mathbf{u}) ,  \textit{rot}^{i-1}(\mathbf{u})\rangle+\sum_{i=1}^n \sum_{\substack{j=1 \\ j\neq i}}^n x_i x_j \langle \textit{rot}^{i-1}(\mathbf{u}), \textit{rot}^{j-1}(\mathbf{u}) \rangle \\
&=\sum_{i=1}^n x_i^2 \Vert \textit{rot}^{i-1}(\mathbf{u}) \Vert^2+ \sum_{r=1}^{n-1}\smashoperator[r]{\sum_{\substack{i,j\in I_n \\ \lvert i-j\rvert=r}}} x_i x_j \langle \textit{rot}^{i-1}(\mathbf{u}), \textit{rot}^{j-1}(\mathbf{u}) \rangle \\
&=\Vert \mathbf{u} \Vert^2 \sum_{i=1}^n x_i^2+  \tau_n 2\smashoperator[lr]{\sum_{\substack{i,j\in I_n \\ i<j \\ j-i=\frac{n}{2}}}} x_i x_j \langle \textit{rot}^{i-1}(\mathbf{u}), \textit{rot}^{j-1}(\mathbf{u}) \rangle +\\ &+ 2\sum_{\substack{r=1 \\ r\neq \frac{n}{2}}}^{n-1}\smashoperator[r]{\sum_{\substack{i,j\in I_n \\ i<j \\ j-i=r}}} x_i x_j \langle \textit{rot}^{i-1}(\mathbf{u}), \textit{rot}^{j-1}(\mathbf{u}) \rangle\\
&=(\rho_1^2+...+\rho_n^2)\sum_{i=1}^n x_i^2+ \tau_n 2\smashoperator[lr]{\sum_{\substack{i,j\in I_n \\ i<j \\ j-i=\frac{n}{2}}}} x_i x_j \Big( 2\Pn{\tfrac{n}{2}}{u}\Big) +\\ &+ 2\sum_{\substack{r=1 \\ r\neq \frac{n}{2}}}^{n-1}\smashoperator[r]{\sum_{\substack{i,j\in I_n \\ i<j \\ j-i=r}}} x_i x_j \Pn{r}{u} =(a^2-2b)\sum_{i=1}^n x_i^2+\\ &+  \tau_n 4\Pn{\tfrac{n}{2}}{u}\smashoperator[lr]{\sum_{\substack{i,j\in I_n \\ i<j \\ j-i=\frac{n}{2}}}} x_i x_j + 2\sum_{\substack{r=1 \\ r\neq \frac{n}{2}}}^{n-1} \vast( \Pn{r}{u} \smashoperator[r]{\sum_{\substack{i,j\in I_n \\ i<j \\ j-i=r}}} x_i x_j \vast)\\
&=(a^2-2b)\sum_{i=1}^n x_i^2+ \tau_n 4\Pn{\tfrac{n}{2}}{u}\, \Pn{\tfrac{n}{2}}{x} + 2\sum_{\substack{r=1 \\ r\neq \frac{n}{2}}}^{n-1} \Pn{r}{u} \, \pn{r}{x}.
\end{align*}

Note that, if $r\neq\frac{n}{2}$, then
\[
\Pn{r}{x}=\pn{r}{x}+\pn{n-r}{x}=\pn{n-(n-r)}{x}+\pn{n-r}{x}=\Pn{n-r}{x}.
\]  
Hence, if $n$ is even,
\begin{align*}
2\sum_{\substack{r=1 \\ r\neq\frac{n}{2}}}^{n-1}\Pn{r}{u} \, \pn{r}{x}
=& \, 2\Big( \Pn{1}{u} \, \pn{1}{x}+\Pn{2}{u} \, \pn{2}{x}+...+ \\ &+ \Pn{\tfrac{n}{2}-1}{u} \, \pn{ \tfrac{n}{2}-1}{x} \, 
+ \\ &+ \Pn{\tfrac{n}{2}+1}{u} \pn{\tfrac{n}{2}+1}{x}+...+ \\ &+\Pn{n-1}{u} \, \pn{n-1}{x} \Big)\\
=& \, 2\Big( \Pn{1}{u} \, \pn{1}{x}+\Pn{n-1}{u} \, \pn{n-1}{x}+\\
&+\Pn{2}{u} \, \pn{2}{x}+\Pn{n-2}{u} \, \pn{n-2}{x}+\\
&+...+\Pn{\tfrac{n}{2}-1}{u} \, \pn{\tfrac{n}{2}-1}{x}+ \\&+ \Pn{\tfrac{n}{2}+1}{u} \, \pn{\tfrac{n}{2}+1}{x} \Big)\\
=& \, 2\sum_{r=1}^{\frac{n}{2}-1}\Pn{r}{u} \,(\pn{r}{x}+\pn{n-r}{x})\\
=& \, 2\sum_{r=1}^{\frac{n}{2}-1}\Pn{r}{u} \, \Pn{r}{x}= \, 2\sum_{r=1}^{\frac{n-2}{2}}\Pn{r}{u} \, \Pn{r}{x}.
\end{align*}

While, if $n$ is odd,
\begin{align*}
2\sum_{\substack{r=1 \\ r\neq\frac{n}{2}}}^{n-1}\Pn{r}{u} \, \pn{r}{x}
=& \, 2\big( \Pn{1}{u} \, \pn{1}{x}+\Pn{2}{u} \, \pn{2}{x}+...+ \\ &+ \Pn{n-1}{u} \, \pn{n-1}{x}\big)\\
=& \, 2\Big( \Pn{1}{u} \, \pn{1}{x}+\Pn{n-1}{u} \, \pn{n-1}{x}+\\
&+\Pn{2}{u} \, \pn{2}{x}+\Pn{n-2}{u} \, \pn{n-2}{x}+\\
&+...+\Pn{\tfrac{n-1}{2}}{u} \, \pn{\tfrac{n-1}{2}}{x}+\\&+  \Pn{\tfrac{n-1}{2}+1}{u} \, \pn{\tfrac{n-1}{2}+1}{x} \Big)\\
=& \, 2\sum_{r=1}^{\frac{n-1}{2}}\Pn{r}{u} \, (\pn{r}{x}+\pn{n-r}{x})\\
=& \, 2\sum_{r=1}^{\frac{n-1}{2}}\Pn{r}{u}\, \Pn{r}{x}.
\end{align*}
Since
\[
\bigg\lfloor \frac{n-1}{2}\bigg\rfloor=
\begin{cases}
{\frac{n-1}{2}}&{\text{if}}\ n {\text{ is odd}}\\
{\frac{n-2}{2}}&{\text{if}}\ n {\text{ is even}},
\end{cases}
\]
then
\[
2\sum_{\substack{r=1 \\ r\neq\frac{n}{2}}}^{n-1}\Pn{r}{u} \, \pn{r}{x}=2\sum_{r=1}^{\lfloor \frac{n-1}{2}\rfloor}\Pn{r}{u} \, \Pn{r}{x}.
\]
Therefore,
\[
\Vert \textit{\textbf{w}} \Vert^2=(a^2-2b)\sum_{i=1}^n x_i^2+ \tau_n 4\Pn{\tfrac{n}{2}}{u}\, \Pn{\tfrac{n}{2}}{x} +2\sum_{r=1}^{\lfloor \frac{n-1}{2}\rfloor} \Pn{r}{u} \, \Pn{r}{x},
\]which proves the theorem.
\end{proof}
\end{theorem}

To make it easier to calculate the minimum norm, our strategy is to make all $\Pn{r}{u}$ zero except for at most a single $r_0\in \{1,2,...,\lfloor n/2 \rfloor \}.$  We want therefore solutions for the system 
\[ \Pn{1}{u}=...=\Pn{r_0-1}{u}=\Pn{r_0+1}{u}=...=\Pn{\lfloor n/2 \rfloor}{u}=0.\]

This system is equivalent to $\langle \mathbf{u},\textit{rot}^{\,r}(\mathbf{u})\rangle=0$ for each $r\in \{1,2,...,r_0-1,r_0+1,...,\lfloor n/2 \rfloor\}$. So geometrically, we want a vector $\mathbf{u}$ that is orthogonal with its rotational shifts, except for at most $\textit{rot}^{\,r_0}(\mathbf{u})$.


 
 This way, we will be able to have the norm of an arbitrary vector $\mathbf{x} G_{\mathbf{u}} \in  \Lambda_\mathbf{u}$ in terms of $a$ and $b$.
 
 
 It is not always simple to obtain an analytic solution for the system. In higher dimensions, it is expected that, from the computational point of view, numerical solutions can be more easily obtained.

\begin{corollary}\label{cons1}
Let $n\geq 2$, $\rho_1,...,\rho_n\in\mathbb{R}$ and $\mathbf{u}=(\rho_1,...,\rho_n)$ such that $\det G_{\mathbf{u}}\neq 0$. If $\Pn{1}{u}=...=\Pn{r_0-1}{u}=\Pn{r_0+1}{u}=...=\Pn{\lfloor n/2 \rfloor}{u}=0$ for some $r_0\in\{ 1,2,...,\lfloor n/2 \rfloor\}$, then for each $
\textit{\textbf{w}}=  \sum_{i=1}^n x_i \textit{rot}^{i-1}(\mathbf{u}) \in\Lambda_\mathbf{u}$,
\[
\Vert \textit{\textbf{w}} \Vert^2=
\begin{cases}
{(a^2-2b) \displaystyle\sum_{i=1}^n x_i^2+4b\Pn{r_0}{x},} &{\text{if}}\ n\ {\text{is even and}}\ r_0=\frac{n}{2}\\
{(a^2-2b) \displaystyle\sum_{i=1}^n x_i^2+2b\Pn{r_0}{x},}&{\text{ otherwise,}}
\end{cases}
\]
where $a,b\in\mathbb{R}$ are the coefficients multiplying $t^{n-1}$ and $t^{n-2}$ respectively in $f(t)= \prod_{i=1}^{n}(t-\rho_i)$.
\end{corollary}


\section{Finding the Determinant of the Generating Matrix}



Within the hypothesis of Corollary \ref{cons1}, we can simplify the expression for $\det G_\mathbf{u}$, which is the main goal of this section, and will be key to compute the center density of $\Lambda_{\textbf{u}}$ later on.
We shall nevertheless recall the complex element $\zeta_n = \cos (2\pi / n) + \sqrt{-1} \, \sin (2\pi / n)$, which is a primitive $n$-th root of unity.

\begin{theorem}\label{tipo1det}
Let $n\geq 2$, $\rho_1,...,\rho_n\in\mathbb{R}$ and $\mathbf{u}=(\rho_1,...,\rho_n)$. If $\Pn{1}{u}=...=\Pn{r_0-1}{u}=\Pn{r_0+1}{u}=...=\Pn{\lfloor n/2 \rfloor}{u}=0$ for some $r_0\in\{ 1,2,...,\lfloor n/2 \rfloor\}$, then
\[
\det G_{\mathbf{u}}=
\begin{cases}
{-a  \displaystyle\prod_{j=1}^{\frac{n-1}{2}}\big( a^2-2b+b\big( \zeta_n^{r_0j}+\zeta_n^{-r_0j} \big) \big),}\, {\text{if}}\ n\ {\text{is odd}}\\
{\pm a^2  \displaystyle\prod_{j=1}^{\frac{n-2}{2}}\big( a^2-2b+b\big( \zeta_n^{r_0j}+\zeta_n^{-r_0j} \big) \big),}\,  {\text{if}}\ n\ {\text{is even and}} \\ \qquad \qquad  \qquad \qquad \qquad \qquad \qquad \quad  \ r_0\ {\text{is even}} \\
{\pm a\sqrt{a^2-4b} \displaystyle\prod_{j=1}^{\frac{n-2}{2}}\big( a^2-2b+b\big( \zeta_n^{r_0j}+\zeta_n^{-r_0j} \big) \big),}\, {\text{if}}\ n\ \text{is even} \\ \quad \quad \qquad \qquad \qquad \qquad \qquad \qquad \qquad \qquad  \text{and}\ r_0\ {\text{is odd}.}
\end{cases}
\]
\begin{proof}
Since $G_{\mathbf{u}}$ is circulant, its eigenvalues are of the form $\lambda_j=\rho_1+\rho_2 \zeta_n^j+...+\rho_n \zeta_n^{(n-1)j}$, $j=0,1,...,n-1$.\

Suppose for now that $n$ is odd.\

It is known that the determinant of a matrix is the product of its eigenvalues, that is,
\begin{align*}
\det G_{\mathbf{u}} &= \prod_{j=0}^{n-1} (\rho_1+\rho_2 \zeta_n^j+...+\rho_n \zeta_n^{(n-1)j})=(\rho_1+\rho_2+...+\rho_n) \prod_{j=1}^{n-1} (\rho_1+\\ &+ \rho_2 \zeta_n^j+...+\rho_n \zeta_n^{(n-1)j})=-a\prod_{j=1}^{n-1} (\rho_1+\rho_2 \zeta_n^j+...+\rho_n \zeta_n^{(n-1)j}).
\end{align*}
Now,
\begin{align*}
\prod_{j=1}^{n-1} (\rho_1+\rho_2 \zeta_n^j+...+\rho_n \zeta_n^{(n-1)j}) = &\prod_{j=1}^{\frac{n-1}{2}} [(\rho_1+\rho_2 \zeta_n^j+...+\rho_n \zeta_n^{(n-1)j}) \\ & (\rho_1+\rho_2 \zeta_n^{n-j}+...+\rho_n \zeta_n^{(n-1)(n-j)})]\\
=& \prod_{j=1}^{\frac{n-1}{2}} [(\rho_1+\rho_2 \zeta_n^j+...+\rho_n \zeta_n^{(n-1)j}) \\ & (\rho_1+\rho_2 \zeta_n^{-j}+...+\rho_n \zeta_n^{-(n-1)j})].
\end{align*}

Note that each term of the product above is of the form
$$
(\rho_1^2+...+\rho_n^2)+\Pn{1}{u} \, (\zeta_n^j+\zeta_n^{-j})+\Pn{2}{u} \, (\zeta_n^{2j}+\zeta_n^{-2j})+ ...+$$  $$ + \Pn{\tfrac{n-1}{2}}{u} \, \Big(\zeta_n^{\frac{n-1}{2}j}+\zeta_n^{-\frac{n-1}{2}j}\Big).
$$
Hence, since $\Pn{1}{u}=...=\Pn{r_0-1}{u}=\Pn{r_0+1}{u}=...=\linebreak \mathcal{P}_n((n-1)/2)=0$, we have
\begin{align*}
\det G_{\mathbf{u}} &=-a\prod_{j=1}^{\frac{n-1}{2}}\big( a^2-2b+\Pn{r_0}{u} \, \big(\zeta_n^{r_0 j}+\zeta_n^{-r_0 j}\big) \big) \\
&=-a\prod_{j=1}^{\frac{n-1}{2}}\big(a^2-2b+b\big(\zeta_n^{r_0 j}+\zeta_n^{-r_0 j}\big)\big).
\end{align*}
On the other hand, suppose that $n$ is even. Then,
\begin{align*}
\det G_{\mathbf{u}} =& \prod_{j=0}^{n-1} (\rho_1+\rho_2 \zeta_n^j+\rho_3\zeta_n^{2j}+...+\rho_n \zeta_n^{(n-1)j})
=(\rho_1+\rho_2+...+\rho_n)\\ & (\rho_1+\rho_2 \zeta_n^{\frac{n}{2}}+\rho_3+...+\rho_{n-1}+\rho_n\zeta_n^{\frac{n}{2}})\prod_{\substack{j=1 \\ j\neq \frac{n}{2}}}^{n-1} (\rho_1+\rho_2 \zeta_n^j+...+\\ &+\rho_n \zeta_n^{(n-1)j})=-a(\rho_1-\rho_2+\rho_3-...+\rho_{n-1}-\rho_n) \prod_{\substack{j=1 \\ j\neq \frac{n}{2}}}^{n-1} (\rho_1+\\ &+ \rho_2 \zeta_n^j+...+\rho_n \zeta_n^{(n-1)j}).
\end{align*}
If $r_0$ is even, let us show that $\rho_1-\rho_2+...+\rho_{n-1}-\rho_n=\pm a$. If $\rho_2+\rho_4+...+\rho_n=0$, then $\rho_1-\rho_2+...+\rho_{n-1}+\rho_n=\rho_1+\rho_3+...+\rho_{n-1}=-a$. If $\rho_2+\rho_4+...+\rho_n\neq 0$, notice that
\[\begin{array}{ll}
(\rho_2+\rho_4+...+\rho_n)(-a) =& (\rho_2+\rho_4+...+\rho_n)[(\rho_2+\rho_4+...+\rho_n)+\\ 
& (\rho_1+\rho_3+...+\rho_{n-1})]\\
&= (\rho_2+\rho_4+...+\rho_n)^2+\smashoperator{\sum_{\substack{i,j\in I_n \\ i\text{ even} \\ j\text{ odd}}}} \rho_i \rho_j\\
&= (\rho_2+\rho_4+...+\rho_n)^2+ \smashoperator{\sum_{\substack{1\leq r \leq \frac{n}{2} \\ r\text{ odd}}}} \Pn{r}{u} \\
&=(\rho_2+\rho_4+...+\rho_n)^2,
\end{array}\]
where we used the fact that $\Pn{r}{u} = 0$ whenever $r$ is odd, since $r_0$ is even.\
Hence, $-a=\rho_2+\rho_4+...+\rho_n$. 

We have also used the fact that if $n$ is even, then $n-r$ has the same parity of $r$, for each $r\in\{1,2,...,n/2 \}$. Consequently, each term $\rho_i \rho_j$ of the sum $\Pn{r}{u}$ has indexes $i$ and $j$ of same parity when $r$ is even, and distinct parities if $r$ is odd.

Now, $-a=\rho_1+...+\rho_n$, so $\rho_1+\rho_3+...+\rho_{n-1}=0$. Hence $\rho_1-\rho_2+...+\rho_{n-1}-\rho_n=-(\rho_2+\rho_4+...+\rho_n)=a$.\

Thus, if $r_0$ is even,
\begin{align*}
\det G_{\mathbf{u}}=&\pm a^2\prod_{\substack{j=1 \\ j\neq \frac{n}{2}}}^{n-1} (\rho_1+\rho_2 \zeta_n^j+...+\rho_n \zeta_n^{(n-1)j})\\
=&\pm a^2\prod_{j=1}^{\frac{n}{2}-1}[(\rho_1+\rho_2 \zeta_n^j+...+\rho_n \zeta_n^{(n-1)j})\\ & (\rho_1+\rho_2 \zeta_n^{-j}+...+\rho_n \zeta_n^{-(n-1)j})]\\
=&\pm a^2\prod_{j=1}^{\frac{n-2}{2}}[(\rho_1+\rho_2 \zeta_n^j+...+\rho_n \zeta_n^{(n-1)j})\\ & (\rho_1+\rho_2 \zeta_n^{-j}+...+\rho_n \zeta_n^{-(n-1)j})].
\end{align*}


Once more, each term of the product above is of the form
\[
(\rho_1^2+...+\rho_n^2)+\mathcal{P}_n(1) (\zeta_n^j+\zeta_n^{-j})+\mathcal{P}_n(2)(\zeta_n^{2j}+\zeta_n^{-2j})+ ...+\mathcal{P}_n(\tfrac{n}{2})\Big(\zeta_n^{\frac{n}{2}j}+\zeta_n^{-\frac{n}{2}j}\Big).
\]

Hence, since $\Pn{1}{u}=...=\Pn{r_0-1}{u}=\Pn{r_0+1}{u}=...=\Pn{n/2}{u}=0$, we have
\begin{align*}
\det G_{\mathbf{u}} &=\pm a^2\prod_{j=1}^{\frac{n-2}{2}}\big(a^2-2b+\Pn{r_0}{u} \, \big(\zeta_n^{r_0 j}+\zeta_n^{-r_0 j}\big)\big)\\
&=\pm a^2\prod_{j=1}^{\frac{n-2}{2}}\big(a^2-2b+b\big(\zeta_n^{r_0 j}+\zeta_n^{-r_0 j}\big)\big).
\end{align*}

If $r_0$ on the other hand is odd, then $\rho_1-\rho_2+...+\rho_{n-1}-\rho_n=\pm \sqrt{a^2-4b}$. Indeed,

\noindent $(\rho_1-\rho_2+\rho_3-...+\rho_{n-1}-\rho_n)^2 =  [(\rho_1+\rho_3+...+\rho_{n-1})-(\rho_2+\rho_4+ \\  ...+\rho_n)]^2 
= (\rho_1+\rho_3+...+\rho_{n-1})^2+(\rho_2+\rho_4+ ...+\rho_n)^2 -2(\rho_1+\rho_3+...+\rho_{n-1}) (\rho_2+\rho_4+...+\rho_n)
=  (\rho_1^2+\rho_2^2+...+\rho_n^2)+2\smashoperator{\sum_{\substack{i,j\in I_n \\ i,j\text{ odd} \\ i\neq j}}} \rho_i\rho_j+ 2\smashoperator{\sum_{\substack{i,j\in I_n \\ i,j\text{ even} \\ i\neq j}}} \rho_i\rho_j-2\smashoperator{\sum_{\substack{i,j\in I_n \\ i\text{ even} \\ j\text{ odd}}}} \rho_i\rho_j
=  (a^2-2b)+2\vast( b-\smashoperator{\sum_{\substack{i,j\in I_n \\ i\text{ even} \\ j\text{ odd}}}} \rho_i\rho_j \vast)-  2\smashoperator{\sum_{\substack{i,j\in I_n \\ i\text{ even}\\ j\text{ odd}}}} \rho_i\rho_j
= \, a^2-4\smashoperator{\sum_{\substack{i,j\in I_n \\ i\text{ even} \\ j\text{ odd}}}} \rho_i\rho_j
= \, a^2-4\sum_{\substack{1\leq r\leq \frac{n}{2} \\ r\text{ odd}}} \Pn{r}{u} 
= a^2-4\Pn{r_0}{u}= \, a^2-4b.
$

 Thus, if $r_0$ is odd,
 \begin{align*}
 \det G_{\mathbf{u}} &= \pm a \sqrt{a^2-4b} \prod_{j=1}^{\frac{n-2}{2}}\big(a^2-2b+\Pn{r_0}{u} \, \big(\zeta_n^{r_0 j}+\zeta_n^{-r_0 j}\big)\big)\\
 &=\pm a \sqrt{a^2-4b} \prod_{j=1}^{\frac{n-2}{2}}\big(a^2-2b+b\big(\zeta_n^{r_0 j}+\zeta_n^{-r_0 j}\big)\big),
 \end{align*}
 which proves the theorem.
\end{proof}
\end{theorem}



  \section{Calculating the Center Density} 
 
In the Corollary \ref{cons1} we establish two expressions for $\Vert \textit{\textbf{w}} \Vert^2.$  Let's analyze the density in each case.
 We shall focus initially in the particular case $r_0 \neq \tfrac{n}{2}$.
 
\subsection{A First Approach to Simplify the Center Density}

If $r_0 \neq \tfrac{n}{2}$ then $\Vert  \textit{\textbf{w}} \Vert^2=(a^2-2b) \sum_{i=1}^n x_i^2+2b\Pn{r_0}{x}$. However, one needs to proceed with caution, because solutions for $\Pn{1}{u}=...=\Pn{r_0-1}{u}=\Pn{r_0+1}{u}=...=\Pn{\lfloor n/2\rfloor}{u}=0$ may lead to $\det G_{\textbf{u}}=0$. Let $D$ be the quadratic form over $\mathbb{Z}$ given by
\[
D\mathbf{x}=(a^2-2b) \sum_{i=1}^n x_i^2+2b\Pn{r_0}{x}.
\]
One may verify that $\det G_{\textbf{u}} \neq 0$ if and only if  $D$ is positive definite, since $D\mathbf{x}=\Vert \textit{\textbf{w}} \Vert^2 =\Vert \mathbf{x} G_{\textbf{u}} \Vert^2$.

Within this context, the next theorem provides a sufficient condition for $\det G_\mathbf{u}\neq 0$. We recall the notation for the greatest common divisor between two numbers $n,m \in \mathbb{N}$ as $(m,n)=\gcd(m,n)$, which shall be used from now on.

\begin{theorem}\label{detdif01}
Let $n\geq 2$, $\rho_1,...,\rho_n\in\mathbb{R}$ and $\mathbf{u}=(\rho_1,...,\rho_n)$ such that $\Pn{1}{u}=...=\Pn{r_0-1}{u}=\Pn{r_0+1}{u}=...=\Pn{\lfloor n/2 \rfloor}{u}=0$ for some $r_0\in\{ 1,2,...,\lfloor (n-1)/2\rfloor\}$.
If $n/(r_0,n)\not\in 2\mathbb{Z}$ and $0\neq a^2\geq 4b$, then $D$ is positive definite.
\begin{proof}~
%
%
For each $\mathbf{x}=(x_1,...,x_n)\in\mathbb{Z}^n$, 
\begin{align*}
D\mathbf{x}=& \, (a^2-2b)\sum_{i=1}^n x_i^2+2b\Pn{r_0}{x}
= \, (a^2-2b)\sum_{i=1}^n x_i^2+2b\smashoperator{\sum_{\substack{i,j\in I_n \\ i<j \\ j-i\in\{r_0,n-r_0\}}}} x_i x_j\\
=& \, \frac{a^2}{4}\smashoperator{\sum_{\substack{i,j\in I_n \\ i<j \\ j-i\in\{r_0,n-r_0\}}}} (x_i+x_j)^2+\frac{a^2-4b}{4}\smashoperator{\sum_{\substack{i,j\in I_n \\ i<j \\ j-i\in\{r_0,n-r_0\}}}} (x_i-x_j)^2.
\end{align*}


If $n/(r_0,n)\not\in 2\mathbb{Z}$, then $\mathbf{x}\neq 0,$ which implies $(x_i+ x_j)^2\geq 1$ for some pair $(i,j)\in\{ (i,j)\in I_n \times I_n \colon i<j, \ j-i\in\{r_0,n-r_0\} \}$.\

Indeed, notice that $\forall i\in\mathbb{N}$, $\exists ! \,j \in I_n$ such that $i\equiv j \, (\text{mod} \,  n)$. 
 Define
\begin{align*}
\varphi \colon \mathbb{N} &\to I_n \, ,\\
i&\mapsto j
\end{align*}
and let $\mathbf{x}=(x_1,...,x_n)\neq 0$. Without loss of generality, assume that $x_1\neq 0$, since otherwise it suffices to rotate $\mathbf{x}$ a convenient amount of times.\

Suppose that $(x_i+x_j)^2=0$ for each $(i,j)\in\{ (i,j)\in I_n\times I_n \colon i<j, \ j-i\in\{r_0,n-r_0\} \}$. In particular,
\[
x_1=-x_{\varphi (1+r_0)}=x_{\varphi (1+2r_0)}=...=(-1)^{k_0 - 1}x_{\varphi(1+(k_0-1)r_0)},
\]
where $k_0=\min \{k\in\mathbb{Z}_+^* \colon 1+kr_0\equiv 1 \, (\text{mod} \, n)\}$.\

Hence $k_0$ is even, because $x_1=(-1)^{k_0} x_{\varphi(1+k_0 r_0)}=-x_1$ otherwise, which can only be true if $x_1 = 0$ (contradiction).\

Moreover, $n/(r_0,n) \in \{k\in\mathbb{Z}_+^* \colon 1+kr_0\equiv 1 \, (\text{mod} \, n)\}$, and
$$
1+k_0r_0\equiv 1 \, (\text{mod} \, n) \Rightarrow k_0r_0\equiv 0 \, (\text{mod} \, n) \Rightarrow n\mid k_0r_0  \Rightarrow$$ 
$$ \Rightarrow \frac{n}{(r_0,n)} \mid k_0\frac{r_0}{(r_0,n)} \Rightarrow \frac{n}{(r_0,n)}\mid k_0.
$$

Consequently, $k_0=n/(r_0,n)$. Therefore, $n/(r_0,n)\in 2\mathbb{Z}$.\

Thus, if $n/(r_0,n) \not\in 2\mathbb{Z}$ with $a^2 \geq 4b$ and $a\neq 0$, then $D\mathbf{x}\geq a^2/4>0$, that is, $D$ is positive definite.
\end{proof}
\end{theorem}

We will see that the condition $0 \neq a^2=4b$ particularly yields interesting lattices. It is important to note that under the hypothesis of Theorem \ref{detdif01}, $a^2=4b$ is equivalent to $\Vert \mathbf{u} \Vert^2 = 2\Pn{r_0}{u}$, that is $\langle \mathbf{u},\mathbf{u} \rangle=2\langle \mathbf{u},\textit{rot}^{r_0}(\mathbf{u}) \rangle$.

A geometric consequence is that $\textit{rot}^{\,r_0}(\mathbf{u})\in \{ x\in \mathbb{R}^n \colon \langle x,\mathbf{u} \rangle=\tfrac{1}{2} \Vert \mathbf{u} \Vert^2\} \cap \Lambda_\mathbf{u}$, that is, $\textit{rot}^{\,r_0}(\mathbf{u})$ is lattice vector as close to the origin as to $\mathbf{u}$. Hence, if in particular $\mathbf{u}$ is a minimal vector, then so is $\mathbf{u}-\textit{rot}^{\,r_0}(\mathbf{u})$. Therefore, we should expect $\lvert S(\Lambda_\mathbf{u}) \rvert$ to increase.

Let us define the quadratic form $Q_r^{(n)}:\mathbb{Z}^n\rightarrow \mathbb{Z}$ by $Q_r^{(n)}\mathbf{x}\vcentcolon= \sum_{i=1}^nx_i^2+\Pn{r}{x}$.

\begin{theorem}\label{minimos4b}
Let $n\geq 2$ and $\rho_1,...,\rho_n\in\mathbb{R}$ such that $\Pn{1}{u}=...=\Pn{r_0-1}{u}=\Pn{r_0+1}{u}=...=\Pn{\lfloor n/2\rfloor}{u}=0$ for some $r_0\in\{ 1,2,...,\lfloor (n-1)/2 \rfloor\}$ such that $n/(r_0,n)\not \in 2\mathbb{Z}$. If $0\neq a^2=4b$, then
$$
\lvert\Lambda_\mathbf{u}\rvert=\frac{a^2}{2}
\qquad
\text{and}
\qquad
\lvert S(\Lambda_\mathbf{u})\rvert = \# \{ \mathbf{x}\in\mathbb{Z}^n \colon Q_{r_0}^{(n)}\mathbf{x}=1 \}.
$$
\begin{proof}
Since $a^2=4b$, we have $a^2-2b=2b=a^2/2$. By Theorem \ref{detdif01}, $\forall \mathbf{x}=(x_1,...,x_n)\in\mathbb{Z}^n$,
\[
D\mathbf{x}=\frac{a^2}{2}\bigg(\sum_{i=1}^nx_i^2 + \Pn{r_0}{x} \bigg)\geq 0.
\]
We have an equality above if and only if $\mathbf{x}=\mathbf{0}$. Thus, if $\mathbf{x} \neq \mathbf{0}$, since $a^2/2>0$, we have
\[
Q_{r_0}^{(n)}\mathbf{x}=\sum_{i=1}^nx_i^2 + \Pn{r_0}{x}\geq 1,
\]
for $x_1,...,x_n\in\mathbb{Z}$.
Now, notice that
\begin{equation} \label{quanti} 
\mathbf{x}=(x_1,...,x_n)=(1,0,...,0) \Rightarrow Q_{r_0}^{(n)}\mathbf{x}=\sum_{i=1}^nx_i^2 + \Pn{r_0}{x}=1.
\end{equation} 

Hence, $a^2/2$ is a lower bound for $\{ D\mathbf{x} \colon \mathbf{x}\in\mathbb{Z}^n\backslash \{0\} \}$, while \linebreak $D(1,0,...,0)=a^2/2$. Thus, since $\Vert \textit{\textbf{w}} \Vert^2 = D \mathbf{x}$, from (\ref{minimum}) it follows that $\lvert \Lambda_\mathbf{u}\rvert=\frac{a^2}{2}.$
Moreover, from (\ref{qminimum}) and (\ref{quanti}), 
\begin{align*}
 \lvert S(\Lambda_\mathbf{u})\rvert 
 =& \, \# \bigg\{ \mathbf{x}\in\mathbb{Z}^n \backslash \{0\} \colon D\mathbf{x}=\frac{a^2}{2} \bigg\} \\
 =& \, \# \{ \mathbf{x}\in\mathbb{Z}^n \backslash \{0\} \colon Q_{r_0}^{(n)}\mathbf{x}=1 \},
 \end{align*} which proves the theorem. \end{proof}
\end{theorem}

Given $n\geq 2$, we can easily compute $\lvert S(\Lambda_\mathbf{u})\rvert$ using a software \cite{wolfram},\cite{python}, \cite{nonlinear}.
In low dimensions, analytical solutions for $\Pn{1}{u}=...=\Pn{r_0-1}{u}=\Pn{r_0+1}{u}=...=\Pn{\lfloor \tfrac{n}{2}\rfloor}{u}=0$ can be found. For example, if $n=5$ and $r_0=2$, then $\mathbf{u}=(0,\rho_2,0,0,-\rho_2)$ solves the system.\
As an example of a numerical solution, if $n=5$ and $r_0=1$, then $\mathbf{u}=(-1.67072, -1.43312, 0.577383,  -0.0932472, -0.789051)$ solves the system
\[
 \begin{cases} \mathcal{P}_5(2)=0 \\ 0 \neq a^2=4b. \end{cases}
 \]
Now, regarding the kissing number, one may verify using a software 
that $\lvert S(\Lambda_\mathbf{u})\rvert=\# \{ \mathbf{x}\in\mathbb{Z}^5 \backslash \{0\} \colon Q_{2}^{(5)}\mathbf{x}=1 \}=\# \{ \mathbf{x}\in\mathbb{Z}^5 \backslash \{0\} \colon Q_{1}^{(5)}\mathbf{x}=1 \}=40=\kappa (D_5)$. Moreover, by Theorem \ref{tipo1det}, from $a^2=4b$ we obtain $\det G_u=-a^5/16$ and therefore $\delta(\Lambda_\mathbf{u})=1/(8\sqrt{2})=\delta(D_5)$. So any solution for $n=5$, regardless of the $r_0$ chosen, yields a lattice with the properties of $D_n$.\


Although it seems convenient to have $0\neq a^2=4b$, it is not always possible to do so within the condition $\Pn{1}{u}=...=\Pn{r_0-1}{u}=\Pn{r_0+1}{u}=...=\Pn{\lfloor n/2\rfloor}{u}=0$.

\begin{proposition} \label{>4b}
Let $n\geq 2$, $\rho_1,...,\rho_n\in\mathbb{R}$ and $\mathbf{u}=(\rho_1,...,\rho_n)$. If $\Pn{1}{u}=...=\Pn{r_0-1}{u}=\Pn{r_0+1}{u}=...=\Pn{\lfloor n/2\rfloor}{u}=0$ for some $r_0\in\{ 1,2,...,\lfloor (n-1)/2 \rfloor\}$, and $0\neq a^2=4b$, then
\[
\det G_u \neq 0 \iff \dfrac{n}{(r_0,n)}\not\in 2\mathbb{Z}.
\]
\begin{proof}
Suppose that $n/(r_0,n)\in 2\mathbb{Z}$, that is, there exists $ c\in 2\mathbb{Z}$ such that $n=c(r_0,n)$. Thus, $n$ is even. Moreover, $r_0/(r_0,n)$ is odd, since otherwise we would have $2(r_0,n)$, a number greater than $(r_0, n)$, divi-\linebreak ding both $n$ and $r_0$, a contradiction.

Now, since $c$ is even, we can consider the entry
\begin{align*}
\mathbf{x}&=(\underbrace{\underbrace{1,0,0,...,0}_{(r_0,n)\text{ coordinates}},\underbrace{-1,0,0,...,0}_{(r_0,n)\text{ coordinates}},...,\underbrace{1,0,0,...,0}_{(r_0,n)\text{ coordinates}},\underbrace{-1,0,0,...,0}_{(r_0,n)\text{ coordinates}}}_{c \text{ blocks of } (r_0,n) \text{ coordinates}}).
\end{align*}
If we map $\Pn{r_0}{}$ over the above vector, then each coordinate multiplies the next $r_0$-th coordinate. But this means going through $r_0/(r_0,n)$ blocks of $(r_0,n)$ coordinates, that is, an odd number of blocks. Thus, $\Pn{r_0}{x}=-c$. Consequently,
\[
D\mathbf{x}=c(a^2-2b)+2b(-c)=c(a^2-4b)=0.
\]
Therefore, $D$ is not positive definite.\

The converse follows from Theorem \ref{detdif01}.
\end{proof}
\end{proposition}

In particular, if $n$ is even, we cannot choose an odd $r_0$. In fact, one may easily verify that $n$ is not a power of $2$ if, and only if, there exists $r_0 \in \{1,2,...,\lfloor (n-1)/2 \rfloor\}$ such that $n/(r_0,n)\not\in 2\mathbb{Z}$.\

Now, we can attempt to simplify the expressions in Theorem \ref{tipo1det} assuming $0\neq a^2=4b$.

\begin{theorem}\label{formula}
Let $n\geq 2$, $\rho_1,...,\rho_n\in\mathbb{R}$ and $\mathbf{u}=(\rho_1,...,\rho_n)$ such that $\Pn{1}{u}=...=\Pn{r_0-1}{u}=\Pn{r_0+1}{u}=...=\Pn{\lfloor n/2\rfloor}{u}=0$ for some $r_0\in\{ 1,2,...,\lfloor (n-1)/2\rfloor\}$ such that $n/(r_0,n) \not\in 2\mathbb{Z}$. If $a^2=4b$, then
\[
\det G_u=\pm \frac{a^n}{2^{n-(r_0,n)}}.
\]
\begin{proof}
Let us first assume that $n$ is odd.\

By Theorem \ref{tipo1det}, since $a^2=4b$, we have
\begin{align*}
\det G_u &=-a\prod_{j=1}^{\frac{n-1}{2}}\big( a^2-2b+b\big( \zeta_n^{r_0 j}+\zeta_n^{-r_0 j}\big)\big)
=-ab^{\frac{n-1}{2}} \prod_{j=1}^{\frac{n-1}{2}}(\zeta_n^{r_0 j}+\\ &+\zeta_n^{-r_0 j}+2)
=-\frac{a^n}{2^{n-1}} \prod_{j=1}^{\frac{n-1}{2}}(\zeta_n^{r_0 j}+\zeta_n^{-r_0 j}+2).
\end{align*}
Since $n$ is odd, then $(2,n)=1$. Hence,
\begin{align*}
\prod_{j=1}^{\frac{n-1}{2}} (\zeta_n^{r_0 j}+\zeta_n^{-r_0 j}+2)&=\prod_{j=1}^{\frac{n-1}{2}} (\zeta_n^{2r_0 j}+\zeta_n^{-2r_0 j}+2)
=\prod_{j=1}^{\frac{n-1}{2}}(\zeta_n^{r_0 j} +\zeta_n^{-r_0 j})^2\\ &
=\vast[ \prod_{j=1}^{\frac{n-1}{2}} (\zeta_n^{r_0 j}+\zeta_n^{-r_0 j}) \vast]^2
=\prod_{j=1}^{n-1}(\zeta_n^{r_0j}+\zeta_n^{-r_0 j})\\
&= (\underbrace{\zeta_n \zeta_n^2 ... \zeta_n^{n-1}}_{=1})^{r_0} \prod_{j=1}^{n-1}(\zeta_n^{r_0 j}+\zeta_n^{-r_0 j})\\
&=\prod_{j=1}^{n-1} \zeta_n^{r_0 j}(\zeta_n^{r_0 j}+\zeta_n^{-r_0 j})
=\prod_{j=1}^{n-1} (1+\zeta_n^{2r_0 j}) \\
&=\prod_{j=1}^{n-1} (1+\zeta_n^{r_0 j}).
\end{align*}
Now, notice that
\begin{align*}
\zeta_n^{r_0 j}=1 &\Rightarrow n\mid r_0 j \\
&\Rightarrow \frac{n}{(r_0,n)} \mid \frac{r_0}{(r_0,n)}j \\
&\Rightarrow \frac{n}{(r_0,n)} \mid j \\
&\Rightarrow  j\in \bigg\{ \frac{n}{(r_0,n)},\frac{2n}{(r_0,n)},...,\frac{((r_0,n)-1)n}{(r_0,n)} \bigg\}.
\end{align*}
Thus,
\begin{align*}
\prod_{j=1}^{n-1} (1+\zeta_n^{r_0 j}) = 2^{(r_0,n)-1}\vast( \prod_{\substack{1\leq j\leq n-1 \\ \zeta_n^{r_0 j} \neq 1}} (1+\zeta_n^{r_0 j}) \vast).
\end{align*}
Moreover,
\begin{align*}
\vast( \prod_{\substack{1\leq j\leq n-1 \\ \zeta_n^{r_0 j}\neq 1}} (1-\zeta_n^{r_0 j}) \vast)\vast( \prod_{\substack{1\leq j\leq n-1 \\ \zeta_n^{r_0 j}\neq 1}} (1+\zeta_n^{r_0 j}) \vast) &= \prod_{\substack{1\leq j\leq n-1 \\ \zeta_n^{r_0 j}\neq 1}} (1-\zeta_n^{2 r_0 j}) \\ &= \prod_{\substack{1\leq j\leq n-1 \\ \zeta_n^{r_0 j}\neq 1}} (1-\zeta_n^{r_0 j}).
\end{align*}
In the last equality we have used the fact that $\{ \zeta_n^{2r_0 j} \colon \zeta_n^{r_0 j}\neq 1, \, 1\leq j\leq n-1 \}=\{ \zeta_n^{r_0 j} \colon \zeta_n^{r_0 j}\neq 1 , \, 1\leq j\leq n-1\}$. Let us briefly demonstrate. Let $\zeta_n^{2r_0 j}$ be an arbitrary element from the former set. Hence $n \nmid 2j$, because otherwise we would have $n \mid j$ and consequently $\zeta_n^{r_0 j}=1$, which is not true. Now let $l\in \{1,...,n-1\}$ such that  $\overline{2 j}=\overline{l}$. Then $\zeta_n^{2r_0 j}=\zeta_n^{r_0 (2j)}=\zeta_n^{r_0 l}$. For the other inclusion, simply notice that $\zeta_n^{r_0 j}=\zeta_n^{2r_0 l}$, where $l=\frac{j}{2}$ if $j$ is even, and $l=\frac{n+j}{2}$ if $j$ is odd.  

Thus,
\[
\prod_{\substack{1\leq j\leq n-1 \\ \zeta_n^{r_0 j}\neq 1}} (1+\zeta_n^{r_0 j})=1,
\]
and therefore
\[
\det G_u=-\frac{a^n}{2^{n-1}} 2^{(r_0,n)-1}=-\frac{a^n}{2^{n-(r_0,n)}}.
\]

Suppose now that $n$ is even. Once again, by Theorem \ref{tipo1det} and $a^2=4b$, we have
\begin{align*}
\det G_u &=\pm a^2\prod_{j=1}^{\frac{n-2}{2}}\big(a^2-2b+b\big(\zeta_n^{r_0 j}+\zeta_n^{-r_0 j}\big)\big)\\
&=\pm a^2 b^{\frac{n-2}{2}} \prod_{j=1}^{\frac{n-2}{2}}(\zeta_n^{r_0 j}+\zeta_n^{-r_0 j}+2)\\
&=\pm \frac{a^n}{2^{n-2}} \prod_{j=1}^{\frac{n-2}{2}}(\zeta_n^{r_0 j}+\zeta_n^{-r_0 j}+2).
\end{align*}
Let $k=(r_0,n)$. Then $\zeta_n^{r_0}=\zeta_{n/k}^{r_0/k}$. Moreover, since $n/k\not\in 2\mathbb{Z}$, then $(n/k,2)=1$. Thus,
\begin{align*}
\prod_{j=1}^{\frac{n-2}{2}}(\zeta_n^{r_0 j}+\zeta_n^{-r_0 j}+2) &= \prod_{j=1}^{\frac{n-2}{2}}\Big(\zeta_{n/k}^{\frac{r_0 j}{k}}+\zeta_{n/k}^{-\frac{r_0 j}{k}}+2\Big)
= \prod_{j=1}^{\frac{n-2}{2}}\Big(\zeta_{n/k}^{\frac{2r_0 j}{k}}+\zeta_{n/k}^{-\frac{2r_0 j}{k}}+2\Big)\\
&= \prod_{j=1}^{\frac{n-2}{2}}\Big(\zeta_{n/k}^{\frac{r_0 j}{k}}+\zeta_{n/k}^{-\frac{r_0 j}{k}}\Big)^2
= \prod_{j=1}^{\frac{n-2}{2}}(\zeta_n^{r_0j}+ \zeta_n^{-r_0 j})^2\\
&= \vast[ \prod_{j=1}^{\frac{n-2}{2}}(\zeta_n^{r_0j}+\zeta_n^{-r_0 j}) \vast]^2
=\frac{1}{\zeta_n^{\frac{r_0 n}{2}}+\zeta_n^{-\frac{r_0 n}{2}}} \prod_{j=1}^{n-1}(\zeta_n^{r_0j}+\\ &+ \zeta_n^{-r_0 j})
=\frac{1}{2} \prod_{j=1}^{n-1}(\zeta_n^{r_0j}+\zeta_n^{-r_0 j})
= (\underbrace{\zeta_n \zeta_n^2 ... \zeta_n^{n-1}}_{=1})^{r_0}\\ & \,\,  \frac{1}{2} \prod_{j=1}^{n-1}(\zeta_n^{j}+\zeta_n^{-j})
=\frac{1}{2} \prod_{j=1}^{n-1} \zeta_n^{r_0 j}(\zeta_n^{r_0 j} + \zeta_n^{-r_0 j}) \\
& =\frac{1}{2} \prod_{j=1}^{n-1} (1+\zeta_n^{2r_0 j}) 
=\frac{1}{2} \prod_{j=1}^{n-1} \Big(1+\zeta_{n/k}^{\frac{2r_0 j}{k}}\Big) \\
&=\frac{1}{2} \prod_{j=1}^{n-1} \Big(1+\zeta_{n/k}^{\frac{r_0 j}{k}}\Big) =\frac{1}{2} \prod_{j=1}^{n-1} (1+\zeta_n^{r_0 j}) .
\end{align*}
But $ \prod_{j=1}^{n-1} (1+\zeta_n^{r_0 j})=2^{(r_0,n)-1}$, and consequently
\[
\det G_u=\pm \frac{a^n}{2^{n-2}} 2^{(r_0,n)-2}=\pm \frac{a^n}{2^{n-(r_0,n)}},
\] which proves the theorem.
\end{proof}
\end{theorem}

\begin{corollary}\label{formulaCase1}
Let $n\geq 2$, $\rho_1,...,\rho_n\in\mathbb{R}$ and $\mathbf{u}=(\rho_1,...,\rho_n)$ such that $\Pn{1}{u}=...=\Pn{r_0-1}{u}=\Pn{r_0+1}{u}=...=\Pn{\lfloor n/2\rfloor}{u}=0$ for some $r_0\in\{ 1,2,...,\lfloor (n-1)/2\rfloor\}$ such that $n/(r_0,n)\not\in 2\mathbb{Z}$. If $0\neq a^2=4b$, then
$$
\delta (\Lambda_\mathbf{u})=\frac{1}{2^{(r_0,n)+\frac{n}{2}}}.
$$
\end{corollary}

Note that, in particular, if $(r_0,n)=1$, which is only possible if $n$ is odd given the hypothesis of the priveous result, then $\delta(\Lambda_\mathbf{u})=\delta(D_n)$. Moreover, if $n$ is even, then $\delta (\Lambda_\mathbf{u})<\delta (D_n)$. The best center density obtained this way is when $r=r_0$ minimizes $\min \Big\{ (r, n) \colon r\in\{ 1,2,...,\lfloor \frac{n}{2}\rfloor \}, \ n/(r_0,n)\not\in 2\mathbb{Z} \Big\}$, i.e., when $r_0=2^\alpha$, where $\alpha$ is the power of $2$ in the prime factorization of $n$. We are allowed to take $r_0=2^\alpha$ given that $n/(r_0,n)\not\in 2\mathbb{Z}$, because in this case $n$ is not a power of $2$, and therefore $n=2^\alpha \prod_{i\in J}p_i^{\alpha_i}>2^{\alpha+1}$, i.e., $2^\alpha<n/2$. 

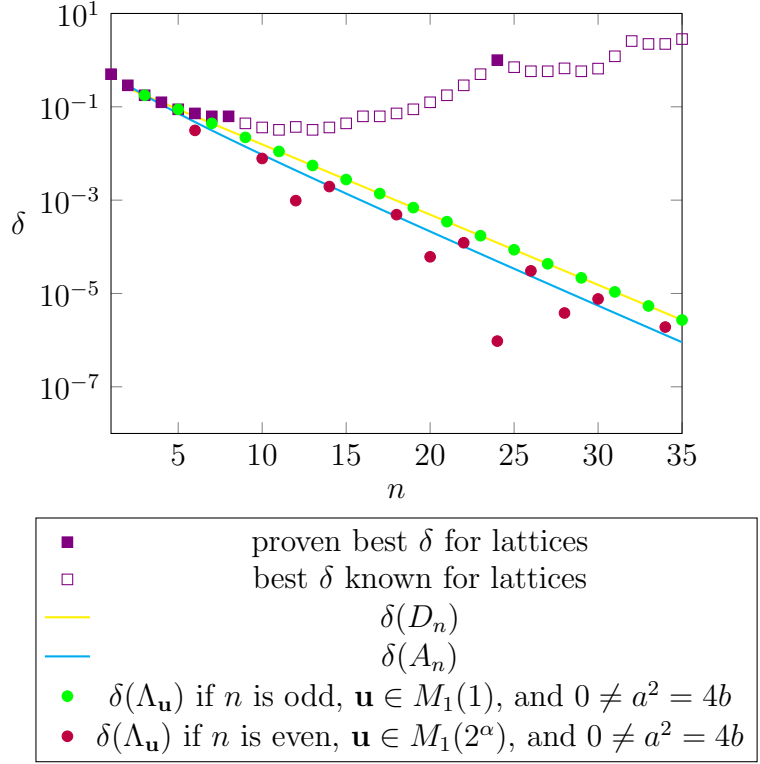
\begin{figure}[!h]
\centering
\begin{tikzpicture}
\begin{semilogyaxis}[height=.2\paperheight,width=.6\linewidth,scale only axis,xmin=1,xmax=35,ymax=10,ymin=10^(-8),xlabel={$n$},ylabel={$\delta$},ylabel style={rotate=-90},legend style={at={(0.5,-0.2)},anchor=north}]


  

  \addplot[only marks, mark=square*,violet] coordinates{(1,0.5) (2,0.28868) (3,0.17678) (4,0.12500) (5,0.08839) (6,0.07217) (7,0.06250) (8,0.06250) (24,1)};
  
  \addplot[only marks, mark=square,violet] coordinates{(9,0.04419) (10,0.03608) (11,0.03208) (12,0.03704) (13,0.03208) (14,0.03608) (15,0.04419) (16,0.06250) (17,0.06250) (18,0.07217) (19,0.08839) (20,0.12500) (21,0.17678) (22,0.28868) (23,0.5) (25,0.70711) (26,0.57735) (27,0.57735) (28,0.66667) (29,0.57735) (30,0.65838) (31,1.20952) (32,2.56578) (33,2.22203) (34,2.22203) (35,2.82843)};

  \addplot[domain=2:35,smooth,thick,yellow]{2^(-1-x/2)};
  
  \addplot[domain=2:35,smooth,thick,cyan]{2^(-x/2)/(x+1)^(1/2)};

  \addplot[domain=3:35,only marks,green,mark=*,samples at={3,5,7,...,35}]{2^(-1-x/2)};

   \addplot[only marks, mark=*, purple] coordinates{(6,1/32) (10,1/128) (12,1/1024) (14,1/512) (18,1/2048) (20,1/16384) (22,1/8192) (24,1/1048576) (26,1/32768) (28,1/262144) (30,1/131072) (34,1/524288)};

   \legend{proven best $\delta$ for lattices \\ best $\delta$ known for lattices \\ $\delta (D_n)$ \\ $\delta (A_n)$\\ $\delta(\Lambda_\mathbf{u})$ if $n$ is odd, $\mathbf{u}\in M_1(1)$, and $0\neq a^2=4b$ \\ $\delta(\Lambda_\mathbf{u})$ if $n$ is even, $\mathbf{u}\in M_1(2^\alpha)$, and $0\neq a^2=4b$ \\}
\end{semilogyaxis}
\end{tikzpicture}
\caption{Center density of $\Lambda_\mathbf{u}$ obtained from Corollary \ref{formulaCase1} if $r_0=2^\alpha$}
\label{densitiesCase1}
\end{figure}

Let $M_1(r_0)=\{\mathbf{u}\in \mathbb{R}^n \colon  \Pn{1}{u}=...=\Pn{r_0-1}{u}=\Pn{r_0+1}{u}=\Pn{\lfloor n/2 \rfloor}{u}\}$ for each $r_0\in\{1,2,...,\lfloor (n-1)/2  \rfloor\}$.
We exhibit in Figure \ref{densitiesCase1} the center densities obtained this way in comparison with lattices such $A_n$ and $D_n$, as well as with the best known center densities.\

\begin{remark}
If $n\in 2\mathbb{Z} \backslash 4\mathbb{Z}$, then $r_0=2.$ We have also $\delta(\Lambda_\mathbf{u})$ within the hypothesis of Theorem \ref{formula}. In this case,
\[
\delta (A_n) > \delta (\Lambda_\mathbf{u}) \Rightarrow \frac{1}{2^{\frac{n}{2}}(n+1)^{\frac{1}{2}}} > \frac{1}{2^{2+\frac{n}{2}}} \Rightarrow 4>(n+1)^{\frac{1}{2}} \Rightarrow 15> n.
\]
Thus, $\delta (\Lambda_\mathbf{u})>\delta(A_n)$ starting with $n=18$.
\end{remark}

\subsection{A Second Approach to Simplify the Center Density}
 By Corollary \ref{cons1}, if $n$ is even, $r_0 = n/2$ and $\Pn{1}{u}=\Pn{2}{u}=...=\Pn{n/2-1}{u}=0$, then $\Vert \textit{\textbf{w}}\Vert^2=(a^2-2b) \sum_{i=1}^n x_i^2+4b \, \Pn{n/2}{x}$.  

Proceeding as in the previous case, we obtain  $\det G_u \neq 0$
 if and only if $a^2>4b.$ Then, we look for conditions between $a^2$ and $b$ that maximize $\delta(\Lambda_{\textbf{u}})$ and we obtain  $a^2=-2b \,(b<0)$ or $a^2=6b\,  (b>0)$. Under these conditions, $\delta(\Lambda_{\textbf{u}})=2^{-n/2}3^{-n/4}.$
 
Note that $\delta(\Lambda_\mathbf{u})=\delta(A_2)$ if $n=2$, the best possible density in this dimension. Starting with $n=4$, however, we have $\delta(\Lambda_\mathbf{u})<\delta(A_n)$, and consequently less convenient densities than in the previous section if $n\in 2\mathbb{Z}\backslash 4\mathbb{Z}$.\



      
    
      
    
      

\section{Conclusion}

\noindent In this paper, we have presented a reasonable expression for the norm of an arbitrary vector in $\Lambda_\mathbf{u}$ and a condition we can assume in order to simplify it. Within this condition, we investigated the hyphotesis $0\neq a^2=4b$, showing that it yields lattices with similar properties to $D_n$.\

The method comes down to solving a system of the form  \vspace{-0.3cm}
\[
\begin{cases}
\Pn{1}{u}=\Pn{2}{u}=...=\Pn{r_0-1}{u}=\Pn{r_0+1}{u}=...\\ =\Pn{\lfloor \tfrac{n}{2}\rfloor}{u}=0 \,\,\,\,\,\,\mbox{and}\\
\Vert \mathbf{u} \Vert^2 = 2\langle \mathbf{u}, \textit{rot}^{r_0} (\mathbf{u}) \rangle,
\end{cases}
\]
where $r_0\in\{1,2,...,\lfloor (n-1)/2 \rfloor\}$ is such that $n/(r_0,n)\not\in 2\mathbb{Z}$.\

The number of equations increases linearly with $n$. Moreover, the optimization of a non-linear system of equations is often difficult to deal with, since comparing float through equality is a source of problem. Thus, in high dimensions it is certainly more convenient to solve a system of the form
\[
(\Vert \mathbf{u} \Vert^2 - 2\langle \mathbf{u}, \textit{rot}^{r_0} (\mathbf{u}) \rangle)^2+\sum_{\substack{r=1 \\ r\neq r_0}}^{\lfloor \frac{n}{2} \rfloor} (\Pn{r}{u})^2<\epsilon
\]
with a sufficiently small $\epsilon>0$. The question is if the solutions for that system yield lattices respecting the results we have presented. We should expect so, since in this case we have that $a^2 \approx 4b$, and $\Vert \sum_{i=1}^n x_i \textit{rot}^{i-1} (\mathbf{u}) \Vert^2$ approximately as in the Corollary \ref{cons1}, for each $\mathbf{x}=(x_1,...,x_n)\in\mathbb{Z}^n$.\


 A single vector $\mathbf{u} \in \mathbb{R}^n$ is needed in order to construct a lattice of the form $\Lambda_\mathbf{u}$, which can be an advantage.
 We obtained in this paper conditions under which $\Lambda_\mathbf{u}$ has the same center density as the $D_n$ lattice in odd dimensions, the best up to the dimension $5$. In even dimensions, our lattices are denser than $A_n$ as long as $n \geq 18$ is not multiple of $4$.
 One may ask themselves what other conditions we may assume over $\mathbf{u}$ in order to obtain dense lattices, or other known classes of lattices.
 
 



\bibliographystyle{IEEEtran}

\begin{thebibliography}{30}

\bibitem{conway} J. H. Conway and N. J. A. Sloane, \emph{Sphere Packings, Lattices and Groups}, 3rd ed. New York, NY: Springer, 1999.

\bibitem{signal1} A. Calderbank and N. Sloane, New trellis codes based on lattices and cosets, \textit{IEEE Transactions on Information Theory}, \textbf{33 (2)} (1987), 177--195. 


\bibitem{cryptography1} A. Joux, J. Stern, Lattice Reduction: A Toolbox for the Cryptanalyst, \textit{J. Cryptology},\textbf{11} (1998), 161--185. 

\bibitem{cryptography2} D. Micciancio and O. Regev,  Lattice-based Cryptography, in \emph{Post-Quantum Cryptography}. D. J. Bernstein, J. Buchmann, E. Dahmen, Ed., Berlin, Heidelberg, Germany: Springer, 2009, 147--191.


\bibitem{2} J. Boutros, E. Viterbo, C. Rastello and J.-C. Belfiore, Good lattice constellations for both Rayleigh fading and Gaussian channels, \textit{IEEE Transactions on Information Theory}, \textbf{42 (2)} (1996),  502--518. 

\bibitem{cohn} H. Cohn and A. Kumar,  Optimality and uniqueness of the Leech lattice among lattices, \textit{Ann. of Math.}, {\textbf{170}} (2009), 1003--1050.





\bibitem{agnaldo} A. J. Ferrari and T. M. R. Souza, Rotated A(n)-lattice codes of full diversity, \textit{Advances In Mathematics Of Communications}, \textbf{16 (3)} (2022), 439--447. 

\bibitem{nphard} I. Dinur, G. Kindler and S. Safra, Approximating-CVP to Within Almost-polynomial Factors is NP-hard, in \textit{Proceedings 39th Annual Symposium on Foundations of Computer Science (Cat. No.98CB36280)}, Palo Alto, CA, USA, (1998), 99--109.

\bibitem{nphard2} JY. Cai, The Complexity of Some Lattice Problems, in \textit{ANTS 2000} in Algorithmic Number Theory, in LNCS, \textbf{1838} (2001), 1--32.


\bibitem{integer1} R. Kannan, Minkowski’s Convex Body Theorem and Integer Programming, \textit{Mathematics of Operations Research}, \textbf{12 (3)} (1988), 415--40. 

\bibitem{integer2} L. Babai, On Lovász’ Lattice Reduction and the Nearest Lattice Point Problem  \textit{Combinatorica}, \textbf{6} (1986),  1--13. 

\bibitem{musin} Musin, O.R.: The problem of twenty-five spheres. Russ. Math. Surv. \textbf{58 (4)} (2003), 744–745.

\bibitem{survey} Boyvalenkov, P., Stefan M. Dodunekov and Oleg R. Musin. Kissing numbers - a survey. \textit{ Computer Scienc}, (2015).

\bibitem{maehara} Maehara, H., Martini, H. Kissing Numbers for Balls with Varying Radii. Graphs and Combinatorics \textbf{38 (183)}, (2022). 

\bibitem{cari} C. Alves, W. L. S. Pinto, A. A. Andrade, 
Well-Rounded Lattices via Polynomials with Real Roots, \textit{International Journal of Applied Mathematics}, \textbf{33 (4)} (2020), 663--672. 


\bibitem{svp} Y. Chuang, C. Fan and Y. Tseng, An Efficient Algorithm for the Shortest Vector Problem, in \textit{IEEE Access}, \textbf{6} (2018), 61478--61487. 

\bibitem{svp2} Z. Sun, C. Gu and Y. Zheng, A Review of Sieve Algorithms in Solving the Shortest Lattice Vector Problem, in \textit{IEEE Access}, \textbf{8} (2020), 190475--190486. 




\bibitem{mici} D. Micciancio, Generalized Compact Knapsacks, Cyclic Lattices, and Efficient One-Way Functions from Worst-Case Complexity Assumptions. \textit{Electron. Colloquium Comput. Complex}, TR04 (2004).

\bibitem{fukshansky} L. Fukshansky and X. Sun, On the Geometry of Cyclic Lattices, \textit{Discrete Comput. Geom.},\textbf{52} (2014), 240--259. 

\bibitem{vinberg} E. B. Vinberg, \textit{A Course in Algebra}, \textbf{56} (2003), Province, RI, USA: AMS, 2003, ch. 3, sec. 3.2.


\bibitem{circulant1} P. J. Davis, \emph{Circulant Matrices}. New York, NY, USA: Wiley-Interscience, 1979.

\bibitem{circulant2} R. M. Gray, Toeplitz and Circulant Matrices: A Review, \textit{Foundations and Trends in Communications and Information Theory}, \textbf{2 (3)} (2006),  155--239. 

\bibitem{wolfram} \textit{Wolfram Research}. (2021). Inc., Mathematica, Version 12.3.1, Champaign, IL.

\bibitem{python} G. Van Rossum and F. L. Drake. (2009). 
\textit{Python 3 Reference Manual}. Scotts Valley, CA: CreateSpace.

\bibitem{nonlinear} P. Bonami, M. Kilinç and J. Linderoth, \textit{Mixed Integer Nonlinear Programming}, 1st ed. New York, NY: Springer, 2012.

 \end{thebibliography}

\end{document}